\documentclass[a4paper,12pt]{article}
\usepackage[dvips]{graphicx}

\begin{document}

\title{Pair creation by a photon in an electric field}
\author{V. N. Baier
and V. M. Katkov\\
Budker Institute of Nuclear Physics,\\ Novosibirsk, 630090, Russia}

\maketitle

\begin{abstract}
The process of pair creation by a photon in a constant and
homogeneous electric field is investigated basing on the
polarization operator in the field. The total probability of the
process is found in a relatively simple form. At high energy the
quasiclassical approximation is valid. The corrections to the
standard quasiclassical approximation (SQA) are calculated. In the
region relatively low photon energies, where SQA is unapplicable,
the new approximation is used. It is shown that in this energy
interval the probability of pair creation by a photon in electric
field exceeds essentially  the corresponding probability in a
magnetic field. This approach is valid at the photon energy much
larger than "vacuum" energy in electric field: $\omega\gg eE/m$. For
smaller photon energies the low energy approximation is developed.
At $\omega\ll eE/m$ the found probability describes the absorption
of soft photon by the particles created by an electric field.

\end{abstract}

\newpage

\section{Introduction}

Pair creation by a photon in an electromagnetic field is the basic
QED reaction which can play the significant role in many processes.

This process was considered first in a magnetic field. Investigation
of pair creation by a photon in a strong magnetic field was started
in 1952 independently by Klepikov and Toll \cite{Kle1, To}. In
Klepikov's paper \cite{Kle2}, which was based on the solution of the
Dirac equation in a constant and homogeneous magnetic field, the
probabilities of radiation from an electron and $e^-e^+$ pair
creation by a photon were obtained for the magnetic field of
arbitrary strength on the mass shell\footnote{We use the system of
units with $\hbar=c=1$ and the metric $ab=a^{\mu}b_{\mu}=a^0b^0-{\bf
a}{\bf b}$} ($k^2=0,~k$ is the 4-momentum of photon). In 1971 Adler
\cite{Ad} calculated the photon polarization operator in the
mentioned magnetic field using the proper-time technique developed
by Schwinger \cite{Schwi} and Batalin and Shabad \cite{Bash}
calculated the photon polarization operator in a constant and
homogeneous electromagnetic field for $k^2\neq 0$ using the Green
function in this field found by Schwinger \cite{Schwi}. In 1975
Strakhovenko and present authors calculated the contribution of a
charged-particles loop with $n$ external photon lines having applied
the proper-time method in a constant and homogeneous electromagnetic
field \cite{BKS}. For $n=2$ the explicit expressions for the
contribution of scalar and spinor particles to the polarization
operator of photon are given. Using this polarization operator the
integral probability of pair creation by a photon in a magnetic
field was analyzed by authors in \cite{BK1}.

The probability of pair creation by a photon in a constant and
homogeneous  electric field in the quasiclassical approximation was
found by Narozhny \cite{Na} using the solution of Dirac equation in
the Sauter potential \cite{Sa}. Nikishov \cite{Ni} obtained the
differential cross section of this process using the solution of
Dirac equation in a constant and homogeneous  electric field.

In the present paper we consider the integral probability of pair
creation in an electric field basing on the polarization operator
\cite{BKS}. In Sec.2 the exact expression for the integral
probability of pair creation by a photon was obtained for the
general case $k^2\neq 0$ starting from the polarization operator in
an electric field. In Sec.3 the standard quasiclassical
approximation (SQA) is outlined for the high-energy photons
$\omega\gg m~(m$ is the electron mass), The corrections to SQA are
calculated. These corrections define also the region applicability
of SQA. In Sec.4 the new approach is developed for relatively low
energies where SQA is not applicable. This approach is based of the
method proposed in \cite{BK1}. The obtained probability is valid in
the wide interval of photon energies and is overlapped with SQA. In
Sec.5 the case of the very low photon energies $\omega\ll m$ is
analyzed. In particular, in the energy region $\omega\leq eE/m$
where the previous approach is unapplicable, the low energy
approximation is developed. In turn the found results have an
overlapping region of applicability with the previous approach. So
we have three overlapping approximations which include all photon
energies. In conclusion we touch upon the problem connected with the
vacuum instability.

\section{Probability of pair creation by a photon: exact theory}

Our analysis is based on the expression for the polarization
operator obtained in \cite{BKS}, see Eqs.(3.19), (3.33). For pure
electric field ({\bf H}=0) this polarization operation can be
written in the diagonal form
\begin{equation}
\Pi^{\mu\nu}=-\sum_{i} \kappa_i\beta_i^{\mu}\beta_i^{\nu},\quad
\beta_i\beta_j=-\delta_{ij}, \quad \beta_ik=0, \quad
\sum_i\beta_i^{\mu}\beta_i^{\nu}=\frac{k^{\mu}k^{\nu}}{k^2}-g^{\mu\nu}.
 \label{d0}
\end{equation}
Here
\begin{eqnarray}
&& \beta_1^{\mu}=\frac{k^2 k_{\perp}^{\mu}+{\bf k}_{\perp}^2
k^{\mu}}{k_{\perp}\sqrt{k^2(\omega^2-k_3^2)}}, \quad
\beta_2^{\mu}=\frac{F^{\mu\nu}k_{\nu}}{E k_{\perp}}, \quad
\beta_3^{\mu}=\frac{F^{\ast\mu\nu}k_{\nu}}{E\sqrt{\omega^2-k_3^2}},
\nonumber \\
&& \kappa_1=\Omega_1(r-q),\quad \kappa_2=\kappa_1+r\Omega_2, \quad
\kappa_3=\kappa_1-q\Omega_3,
\nonumber \\
&& r=\frac{\omega^2-k_3^2}{4m^2}, \quad q=\frac{k_{\perp}^2}{4m^2},
\quad r-q=\frac{k^2}{4m^2}, \label{d1}
\end{eqnarray}
where the axis 3 directed along the electric field {\bf E}, ${\bf
k}_{\perp}{\bf E}=0$, $k_{\perp}=\sqrt{{\bf k}_{\perp}^2}$, $\omega$
is the photon energy, $F^{\mu\nu}$ is the tensor of electromagnetic
field, $F^{\ast\mu\nu}$ is the dual tensor of electromagnetic field,
and
\begin{equation}
\Omega_i=-\frac{\alpha
m^2}{\pi}\int\limits_{-1}^1dv\int\limits_0^{\infty-i0}f_i(v,x)\exp(i\psi(v,x))dx.
\label{d2}
\end{equation}
Here
\begin{eqnarray}
&& f_1(v,x)=\frac{\cosh vx}{\sinh x}-v\frac{\cosh x \sinh
vx}{\sinh^2 x}, \quad f_2(v,x)=2\frac{\cosh x-\cosh
vx}{\sinh^3x}-f_1(v,x),
\nonumber \\
&& f_3(v,x)=(1-v^2)\coth x -f_1(v,x) ,
\nonumber \\
&& \psi(v,x)=\frac{1}{\nu}\left(2r\frac{\cosh x -\cosh vx}{\sinh
x}-x(1+q(1-v^2))\right),\quad \nu=\frac{E}{E_0}.\label{d3}
\end{eqnarray}

Let us note that the integration contour in Eq.(\ref{d2}) is turned
slightly down, and in the function $\Omega_1$ in the integral over
$x$ the subtraction at $\mu=0$ is implied.

It should be noted that the probability of pair creation in an
electric field (see Eqs.(\ref{d0})-(\ref{d3})) can be obtained from
the probability of pair creation in a magnetic field
(Eqs.(2.1)-(2.5) in \cite{BK1}) using the formal substitutions $\mu
\rightarrow i\nu,~x\rightarrow ix,~q\leftrightarrow -r$.

The imaginary part of the polarization operator determines the total
probability of $e^-e^+$ pair creation per unit length $W_i$ by a
photon with a given polarization
\begin{equation}
W=\frac{1}{\omega}~{\rm Im}e^{\mu}e^{\nu\ast}~\Pi_{\mu\nu}.
\label{d4}
\end{equation}
Using the photon polarization vector $e^{\mu}_{i}=\beta^{\mu}_{i}$
we get the expressions for $W_i~(i=1,2,3)$ at $k^2\neq 0$:
\begin{equation}
 W_i=-\frac{{\rm Im}\kappa_i}{\omega}
 \label{d5}
\end{equation}
On the mass shell ($k^2=0$) one has to put $r=q$ in Eq.(\ref{d2}).
In this case $W_1=0$ and only two photon polarizations $i=2$ and
$i=3$ contribute. In this case the total probability of pair
creation averaged over the photon polarizations is
\begin{equation}
 W=\frac{W_2+W_3}{2}.
 \label{d5a}
\end{equation}

The corresponding analysis in a magnetic field ({\bf E}=0) was
performed in \cite{BK1}. There are essential differences between
pair creation process in magnetic and electric field.
\begin{enumerate}

\item The probabilities of pair creation $W_i(r)$ (for each photon
polarization) in a magnetic field  contain the factor $1/\sqrt{g}$
and $g=0$ at each threshold when electrons and positrons are created
on the Landau levels. Because of this the functions $W_i(r)$ have
saw-tooth form. Rather laborious transformations of the imaginary
part of the polarization operator are performed in \cite{BK1} to
obtain the form allowing the direct calculation of the pair creation
probabilities. It's more than difficult to use for this purpose
directly the expression ${\rm Im}(e^{\mu}e^{\nu\ast}~\Pi_{\mu\nu})$.
In an electric field there are no levels and ${\rm
Im}(e^{\mu}e^{\nu\ast}~\Pi_{\mu\nu})$ is a smooth function of $r$
and can be calculated directly using the above equation taking into
account that the integration contour in Eq.(\ref{d2}) is turned
slightly down. The result for $\nu=0.01, 0.001, 0.0001$ is shown in
Fig.1. Here and below in all figures the frame $k_3=0$ is used. In
general case $\omega/2m\rightarrow k_{\perp}/2m=\sqrt{r}$.

\begin{figure}[h]
  \centering
  \includegraphics[width=0.90\textwidth
  ]{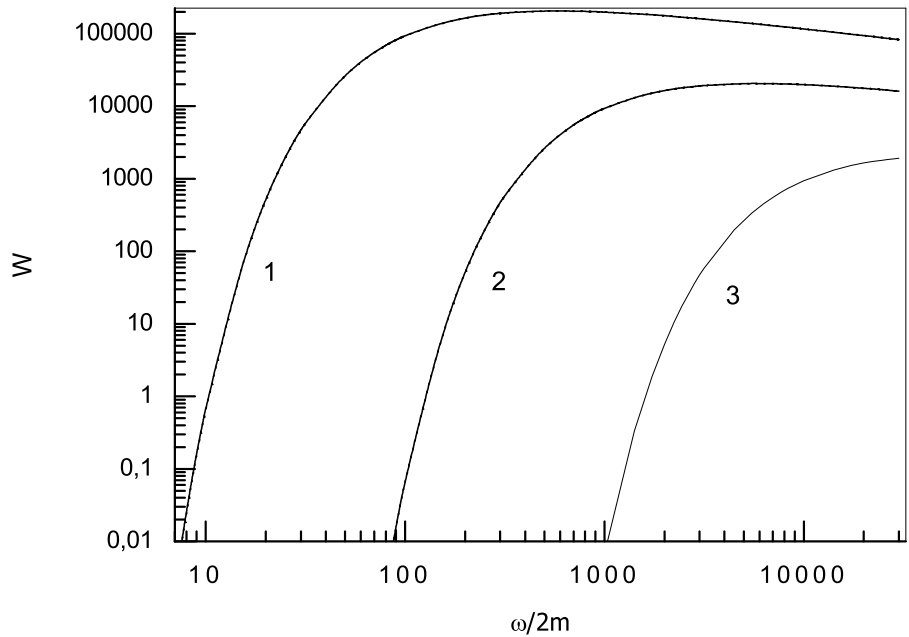}\hspace{0.02\textwidth}
 \caption{The total probability of pair creation by a photon averaged over the
photon polarizations $W(r)$ (in units cm$^{-1}$) Eqs.(\ref{d5}),
(\ref{d5a}) in an electric  field for $\nu=0.01$ (curve 1),
$\nu=0.001$ (curve 2), $\nu=0.0001$(curve 3) vs $\omega/2m$. }
\label{Fig:1}
 \end{figure}

\item
In a magnetic field the pair creation is the result of conversion of
a photon into pair. The constant and homogeneous magnetic field
itself doesn't create a pair. The vacuum is stable. For this reason
Eq.(\ref{d4}) gives not only imaginary part of the photon energy but
else the overall probability of pair creation. The expression for
the polarization operator is valid for field strength $B >
B_0=m^2/e=4.41\cdot 10^{13} G$ (one of possible application of the
theory are processes in magnetars with $B > B_0$). In an electric
field, generally speaking, pairs can be created by field itself,
without a photon presence. The vacuum is unstable. Then
Eq.(\ref{d4}) gives the partial probability of pair creation. In the
limit $E \ll E_0=m^2/e~(\nu\ll 1)$ the instability of vacuum is
negligible. But at $E \sim E_0$ the vacuum pair creation becomes
essential. Even in this situation for $\omega\gg m~(r\gg 1)$ the
photon create high-energy particles, while the electric field create
low-energy particles ($\varepsilon\sim m$) and the pairs are easy
distinguishable.

\item
In a magnetic field there is the threshold ${\rm
Im}(e^{\mu}e^{\nu\ast}~\Pi_{\mu\nu})=0$ for $r < 1$. Strictly
speaking there is no such requirement in an electric field (see
below).

\end{enumerate}

\section{Probability of pair creation by a photon in
quasiclassical approximation}

The standard quasiclassical approximation valid for
ultrarelativistic created particles ($r \gg 1$) can be derived from
Eqs.(\ref{d2}), (\ref{d3}) by expanding the functions  $f_2(v,x)$,
$f_3(v,x),~ \psi(v,x)$ over $x$ powers. Taking into account the
higher powers of $x$ one gets
\begin{eqnarray}
\hspace{-10mm}&&
f_2(v,x)=\frac{1-v^2}{12}\left[-(3+v^2)x+\frac{1}{15}(15-6v^2-v^4)x^3\right]
\nonumber \\
\hspace{-10mm}&&
f_3(v,x)=\frac{1-v^2}{6}\left[(3-v^2)x-\frac{1}{60}(15-2v^2+3v^4)x^3\right]
\nonumber \\
\hspace{-10mm}&&
\psi(v,x)=-\frac{r(1-v^2)^2}{12\nu}\left(x^3-\frac{3-v^2}{30}x^5\right)
-\frac{x}{\nu}.
 \label{c1}
\end{eqnarray}
Here the first terms in the brackets give the known probability of
the process in the standard quasiclassical approximation, while the
second terms are the corrections.Expanding the term with $x^5$ in
$\exp(i\psi(v,x))$ and making substitution $x=\nu t$ one finds
\begin{eqnarray}
\hspace{-15mm}&& {\rm Im}\Omega_i = i\frac{\alpha
m^2\nu}{2\pi}\int\limits_{-1}^{1}dv\int\limits_{-\infty}^{\infty}
g_i(v,t)\exp\left[-i\left(t+\xi\frac{t^3}{3}\right)\right]dx,
\nonumber \\
\hspace{-15mm}&& g_2(v,t)=\frac{1-v^2}{12}\nu
t\left[-(3+v^2)-i\frac{9-v^4}{90}\xi\nu^2t^5+
\frac{\nu^2t^2}{15}(15-6v^2-v^4)\right],
\nonumber \\
\hspace{-15mm}&& g_3(v,t)=\frac{1-v^2}{6}\nu
t\left[(3-v^2)+i\frac{(3-v^2)^2}{90}\xi\nu^2t^5-
\frac{\nu^2t^2}{60}(15-2v^2+3v^4)\right],
 \label{c2}
\end{eqnarray}
where
\begin{equation}
\xi=\frac{(1-v^2)^2\kappa^2}{16},\quad \kappa^2=4r\nu^2. \label{c3}
\end{equation}
We will use the known integrals
\begin{eqnarray}
\hspace{-5mm}&&
\int\limits_{-\infty}^{\infty}\cos\left(t+\xi\frac{t^3}{3}\right)
=\sqrt{3}z K_{1/3}(z),\quad
z=\frac{2}{3\sqrt{\xi}}=\frac{8}{3(1-v^2)\kappa},
\nonumber \\
\hspace{-5mm}&&
\int\limits_{-\infty}^{\infty}t\sin\left(t+\xi\frac{t^3}{3}\right)
=\frac{3\sqrt{3}}{2}z^2 K_{2/3}(z).
 \label{c4}
\end{eqnarray}
Conserving the main(first) terms of functions $g_n(v,t)$ in the
integrals Eq.(\ref{c2}) and taking into account
Eqs.(\ref{c3})-(\ref{c4}), we obtain the probabilities of pair
creation in standard quasiclassical approximation
\begin{equation}
W_n^{(SQA)}=-{\rm Im}\frac{\kappa_n}{\omega} =\frac{\alpha
m^2}{3\sqrt{3}\pi\omega}
\int\limits_{-1}^{1}\frac{s_n}{1-v^2}K_{2/3}(z)dv,\quad
s_2=3+v^2,\quad s_3=2(3-v^2)
 \label{c6}
\end{equation}
Here Eq.(\ref{c6}) coincides with the probability obtained in
Appendix C \cite{BK1} for the case of magnetic field. This is
because in the quasiclassical approximation the expression for the
probability depends on an electromagnetic field via parameter
$\kappa^2=4(r\nu^2+q\mu^2)$ (in the frame where electric and
magnetic fields are parallel).

Below we will see that at lower energy the probabilities of pair
photoproduction are very different in electric and in magnetic
fields.

The probability of pair creation by a photon averaged over photon
polarization Eq.(\ref{d5a}) calculated in this approximation
Eq.(\ref{c6}) coincide with curves found in calculation of exact
expressions Eqs.(\ref{d5})-(\ref{d5a}) given in Fig.1. Near maximum
of the curves the agreement is better than $\sim 10^{-5}$.

The corrections to the standard approximation can be found from the
mentioned Appendix C (see Eqs.(C7)-(C12)) by the substitution $\mu^2
\rightarrow -\nu^2$
\begin{equation}
W_i^{(1)}=-\frac{\alpha m^2\nu^2}{30\sqrt{3}\pi\omega\kappa}
\int\limits_{0}^{1}G_i(v,z)\frac{dv}{1-v^2}, \label{c7}
\end{equation}
where
\begin{eqnarray}
&& G_2(v,z)=(36+4v^2-18z^2)K_{1/3}(z)+(3v^2-57)zK_{2/3}(z),
\nonumber \\
&& G_3(v,z)=-(34+2v^2+36z^2)K_{1/3}(z)+(78-6v^2)zK_{2/3}(z).
 \label{c8}
\end{eqnarray}
The asymptotic at $\kappa \gg 1$ are
\begin{eqnarray}
&& W_i^{(1)}=-\frac{\alpha m^2\nu^2}{30\sqrt{3}\pi\omega\kappa}w_i,
\quad w_2=12 A \kappa^{1/3}-90\pi,
\nonumber \\
&&  w_3=-11 A \kappa^{1/3}+84\pi,\quad
A=3^{1/3}\frac{2}{5}\frac{\Gamma^3(1/3)}{\Gamma(2/3)} = 8.191.. .
\nonumber \\
&& W^{(1)}=\frac{W_1^{(1)}+W_2^{(1)}}{2}=-\frac{\alpha m^2\nu^2}
{60\sqrt{3}\pi\omega\kappa}\left(A \kappa^{1/3}-6\pi+...\right)
\nonumber \\
&& \frac{W^{(1)}}{W^{(SQA)}}=-3^{-7/3}\frac{7}{125}
\frac{\Gamma^4(1/3)}{\Gamma^4(2/3)}\frac{\nu^2}{\kappa^{4/3}}
\left(1-\frac{6\pi}{A\kappa^{1/3}}+...\right)
 \label{c9}
\end{eqnarray}
At $\kappa \ll 1$ one has
\begin{equation}
W_2^{(1)}=\frac{\alpha
m^2\nu^2}{\omega\kappa^2}\frac{2\sqrt{2}}{5\sqrt{3}}
\exp\left(-\frac{8}{3\kappa}\right), \quad
W_3^{(1)}=2W_2^{(1)},\quad
\frac{W^{(1)}}{W^{(SQA)}}=\frac{32\nu^2}{15\kappa^3}.
 \label{c10}
\end{equation}
The curves in Fig.2 characterize the applicability of SQA at
energies which are lower than shown in Fig.1. It is seen that for
small $\nu$ the applicability of SQA is broken. For curves 3 and 4,
where parameter $\kappa \ll 1$, the value $R(10)-1$ agrees with the
last correction in Eq.(\ref{c10}).

\begin{figure}[h]
  \centering
  \includegraphics[width=0.52\textwidth
  ]{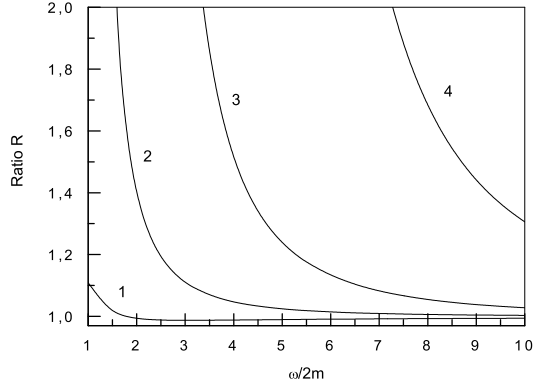}\hspace{0.025\textwidth}
 \caption{The ratio of total probabilities of pair creation by a photon
averaged over the photon polarizations Eq.(\ref{d5a})
$R=W^{(ex)}/W^{(SQA)}$ vs $\omega/2m$, where $W^{(ex)}$ is given by
Eqs.(\ref{d5})-(\ref{d5a}), $W^{(SQA)}$ is given by
Eqs.(\ref{c6}),(\ref{d5a}). The curve 1 is for $\nu=1$, the curve 2
is for $\nu=0.1$, the curve 3 is for $\nu=0.01$, the curve 4 is for
$\nu=0.001$.} \label{Fig:2}
 \end{figure}

\section{Region of low photon energies}

In the field which is weak comparing with the critical field
$E/E_0=\nu \ll 1$ and at relatively low photon energy ($r \leq
\nu^{-2/3}$)  the standard SQA \cite{BK,BKF,BKS1} is nonapplicable.
This follows from the last equality in Eq.(\ref{c10}). In this case,
if the condition $r \gg \nu^2$ is fulfilled, the method of
stationary phase at calculation of the imaginary part of the
integral over x in Eq.(\ref{d2}) can be applied. To this end we
present the imaginary part of $\Omega_i$ in the form
\begin{equation}
{\rm Im}\Omega_i = i\frac{\alpha
m^2}{2\pi}\int\limits_{-1}^{1}dv\int\limits_{-\infty}^{\infty}
f_i(v,x)\exp\left[i\psi(v,x)\right]dx.
\label{b0}
\end{equation}

Granting that the large parameter $1/\nu$ is the common factor in
the phase $\psi(x)$, it doesn't contained in the equation
$\psi'(x)=0$ which defines the stationary phase point $x_0(r \sim
1)\sim 1$. In this case the small values of variable $v$ contribute
to the integral over $v$, so that one can extend the integration
limits to the infinity. So we get
\begin{equation}
{\rm Im}\Omega_i \simeq i\frac{\alpha
m^2}{2\pi}\int\limits_{-\infty}^{\infty}dv\int\limits_{-\infty}^{\infty}
f_i(v=0,x)\exp\left\{-\frac{i}{\nu}
\left[\varphi(x)+v^2\chi(x)\right]\right\}dx,\label{b1}
\end{equation}
where
\begin{equation}
\varphi(x)=2r\tanh\left(\frac{x}{2}\right)+(r+1)x,\quad
\chi(x)=rx\left(-1+\frac{x}{\sinh x}\right).\label{b2}
\end{equation}
From the equation $\varphi'(x)=0$ we find
\begin{equation}
\tanh\left(\frac{x_0}{2}\right)=-\frac{i}{\sqrt{r}}, \quad
\frac{x_0(r)}{2}=-i\frac{a(r)}{2}=-i\arctan\frac{1}{\sqrt{r}}.
\label{b3}
\end{equation}
Substituting these results in the expressions which defines the
integral in Eq.(\ref{b1}) we have
\begin{eqnarray}
\hspace{-10mm}&&
i\varphi(x_0)=2\left((r+1)\arctan\frac{1}{\sqrt{r}}-\sqrt{r}\right)\equiv
b(r),
\nonumber \\
\hspace{-10mm}&& i\varphi''(x_0)=\frac{r+1}{\sqrt{r}}, \quad
i\chi(x_0)=\sqrt{r}a(r)b(r)
\nonumber \\
\hspace{-10mm}&& if_2(v=0, x_0(r))=-\frac{r+1}{2r^{3/2}},\quad
if_3(v=0, x_0(r))=\frac{1}{\sqrt{r}}.
 \label{b4}
\end{eqnarray}
Performing the standard procedure of the method of stationary phase
one obtains for probability of pair creation in an electric field by
a polarized photon
\begin{equation}
W_2^{(th)}=\frac{\alpha m^2\nu}{2\omega}\sqrt{\frac{r+1}{ra(r)b(r)}}
\exp\left(-\frac{b(r)}{\nu}\right),\quad
W_3^{(th)}=\frac{2r}{r+1}\left(1+\frac{\nu}{4\pi
r}\right)W_2^{(th)}, \label{b5}
\end{equation}
where the term $\nu/4\pi r$ in $W_3^{(th)}$ valid at $r\ll 1$
appears as the contribution of the second term in $f_1(v, x)
(\propto v^2)$ in Eq.(\ref{d3}). These probabilities can be found
from the probabilities of pair creation in a magnetic field (see
Eqs.(B3)-(B5) (Appendix B in \cite{BK1} by substitutions
$\mu\rightarrow i\nu,~ r\rightarrow -r,~\sqrt{-r}=-i\sqrt{r}$. At
this substitution $l(r)\rightarrow l(-r)=ia(r)$ and
$\beta(r)\rightarrow \beta(-r)=ib(r)$ and from Eq.(B5) in \cite{BK1}
one obtains Eq.(\ref{b5}) (without the correction term $\nu/4\pi r$,
which was not taken into account in  \cite{BK1}).

\begin{figure}[h]
  \centering
  \includegraphics[width=0.60\textwidth
  ]{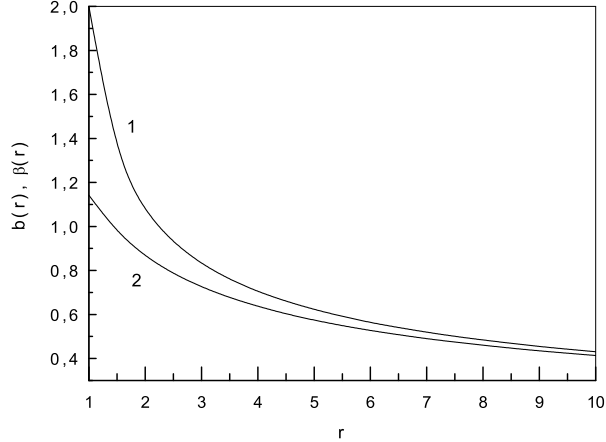}\hspace{0.025\textwidth}
 \caption{The exponent $\beta(r)$ in Eq.(B5) (Appendix B in \cite{BK1}) for
pair creation in a magnetic field (curve 1) and the exponent $b(r)$
in Eq.(\ref{b5}) for pair creation in an electric field (curve 2).}
\label{Fig:3}
 \end{figure}

The comparison of the exponent $b(r)$ with the exponent
$\beta(r)=2\sqrt{r}-(r-1)l(r), l(r)=\ln[(\sqrt{r}+1)/(\sqrt{r}-1)]$
in the mentioned probability of pair creation in a magnetic field is
given in Fig.3. Since the exponent $b(r)$ enters with very large
factor $1/\nu$ and the exponent $\beta(r)$ with the factor $1/\mu$
the probability of pair creation in an electric field is much larger
than the probability of pair creation in a magnetic field at
$\nu=\mu \ll1$.

In spite of the assumption $r \sim 1$ made above, Eq.(\ref{b5}) is
valid also at $r \gg 1$ if the condition $b(r) \gg \nu$ is
fulfilled. This can be traced in the derivation of Eq.(\ref{b5}).
The first two term of the decomposition of the function $b(r)$ over
power of $1/r$ are
\begin{equation}
\frac{b(r)}{\nu}\simeq \frac{4}{3\nu\sqrt{r}}-\frac{4}{15\nu
r^{3/2}}\label{b6}
\end{equation}
It follows from this formula that applicability of Eq.(\ref{b5}) is
limited by the condition $r \ll \nu^{-2}$.  If the second term much
smaller than unity the exponent with it can be expanded. As a result
we have from Eq.(\ref{b5}) at $\nu^{-2/3}\ll r \ll \nu^{-2}$
\begin{equation}
W_3=\frac{\alpha m^2\nu}{2\omega}\sqrt{\frac{3r}{2}}
\exp\left(-\frac{4}{3\nu\sqrt{r}}\right)\left(1+ \frac{4}{15\nu
r^{3/2}}\right), \quad W_2=\frac{1}{2}W_3\label{b7}
\end{equation}
The main term in the above expression coincides with the probability
of pair creation by a photon in the standard quasiclassical theory
at $\kappa=2\nu\sqrt{r} \ll 1$. The correction in Eq.(\ref{b7})
determines the lower energy limit of the standard approach
applicability ($\kappa^3 \gg \nu^2$) and coincide with
Eq.(\ref{c10}). So the overlapping region exists where both the
formulated here and the standard approach for high energy are valid.

\begin{figure}[h]
  \centering
  \includegraphics[width=0.9\textwidth
  ]{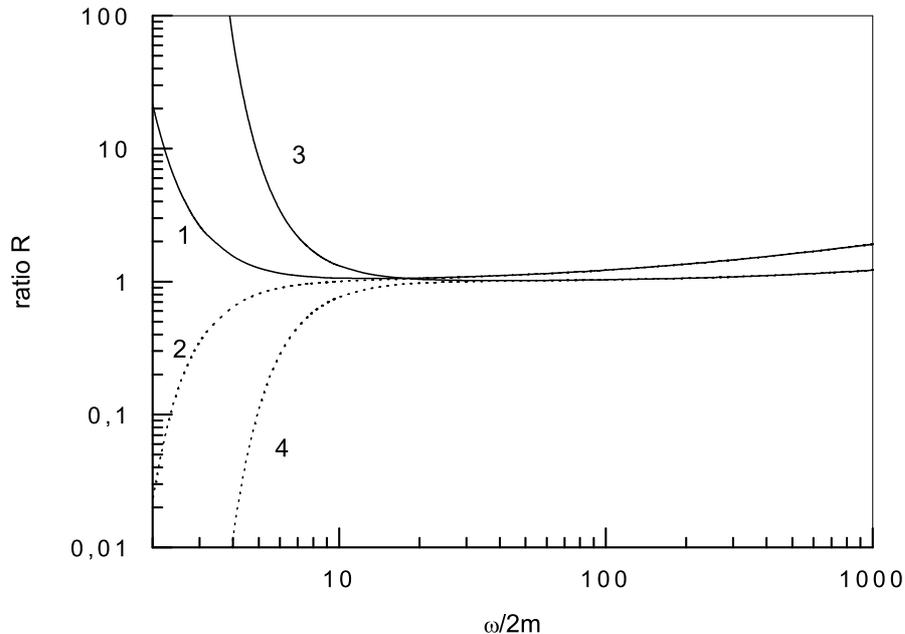}\hspace{0.025\textwidth}
 \caption{The ratio of the total probabilities of pair creation by a photon
averaged over the photon polarizations Eq.(\ref{d5a})
$R=W^{(th)}/W^{(SQA)}$ (see Eqs.(\ref{b5}),(\ref{c6})) is shown for
electric field $\nu=0.01$ (curve 1), magnetic field $\mu=0.01$
(curve 2) and for electric field $\nu=0.001$ (curve 3) and magnetic
field $\mu=0.001$ (curve 4) vs $\omega/2m$. } \label{Fig:4}
 \end{figure}

In Fig.4 the ratio of the probabilities of pair creation by a photon
averaged over the photon polarizations Eq.(\ref{d5a})
$R=W^{(th)}/W^{(SQA)}$ (see Eqs.(\ref{b5}),(\ref{c6})) are given for
an electric field (curves 1,3) and a magnetic field (see
Eqs.(B3)-(B5) in \cite{BK1}) (curves 2,4). It is seen that the
probabilities in an electric field at low $r$ values are many order
of magnitude higher then in the same magnetic field. The probability
in an electric field aims at the standard quasiclassical one with
$r$ increase from above while the probability in a magnetic field
aims at  the standard quasiclassical value from below. According to
Eq.(\ref{b7}), in the interval $\nu^{-2/3}\ll r \ll \nu^{-2}$  the
ratio $R$ is close to unity for both electric and magnetic fields
and for a given $\nu(\mu)$ the curves have been merged, since in SQA
the exponential form of the probabilities coincide with
Eq.(\ref{b7}) (without the correction term which has opposite sign
for a magnetic fields).  It is valid when $\kappa\ll 1$. At
$\kappa=1~(\sqrt{r}=1/2\nu)$ the value $R\simeq 1.13$ and with
further $r$ increase the value $R$ smoothly grows. At very high
values of $\kappa$ the probability $W^{(SQA)}=\alpha m \nu C
\kappa^{-1/3}, C=0.37961$ (see \cite{BKS}, Eq.(3.60)). When this
equation is valid the value $R\propto r^{1/6}$.

At low photon energy ($\nu^2\ll r\ll \nu^{2/3}$) the probability
Eq.(\ref{b5}) has a form
\begin{equation}
W_2\simeq \frac{\alpha
m^2\nu}{2\pi\omega\sqrt{r}}\left(1+\frac{3\sqrt{r}}{\pi}\right)
\exp\left(-\frac{\pi}{\nu}\left(1+r\right)+\frac{4\sqrt{r}}{\nu}\right),\quad
W_3=\left(2r+\frac{\nu}{2\pi}\right) W_2. \label{b8}
\end{equation}

The curves in Fig.5 characterize the applicability of Eq.(\ref{b5})
for different values of parameter $\nu\ll 1$ in wide interval of
$\omega/2m$. The region of applicability of the probability
$W^{(th)}$ Eq.(\ref{b5}) is extended with $\nu$ decrease.

\begin{figure}[h]
  \centering
  \includegraphics[width=0.70\textwidth
  ]{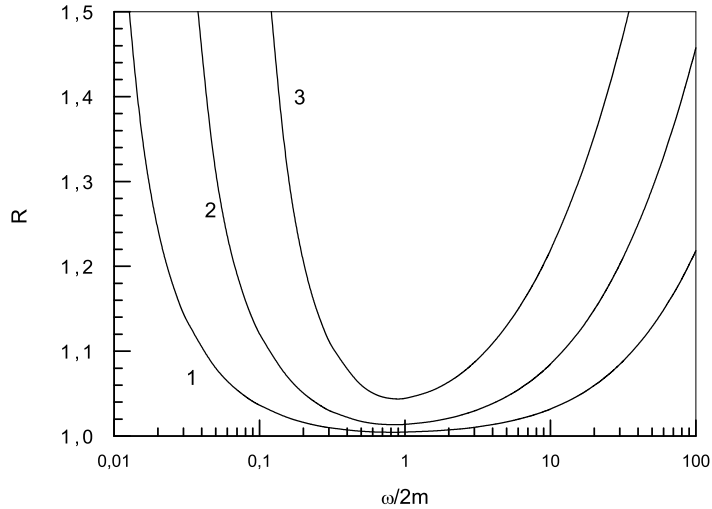}\hspace{0.025\textwidth}
 \caption{The ratio of the total probabilities of pair creation by a photon
averaged over the photon polarizations Eq.(\ref{d5a})
$R=W^{(th)}/W^{(ex)}$ (see Eqs.(\ref{b5}),(\ref{d5})),  for
$\nu=0.01$ (curve 1), for $\nu=0.03$ (curve 2) and for $\nu=0.1$
(curve 3) vs $\omega/2m$. } \label{Fig:5}
 \end{figure}

\section{Approximation at very low photon energy}

In the case $r \ll \nu^2/(1+\nu)$ one can expand the exponent in
Eq.(\ref{b0}) in  powers of the term $\propto r/\nu$. Conserving the
main term $\exp(-ix/\nu)$, independent of variable $v$, performing
the integration over $v$ of the functions $f_i(v,x)$ we find
\begin{equation}
{\rm Im}\Omega_i=-i\frac{\alpha
m^2}{\pi}\int\limits_{-\infty}^{\infty}\varphi_i(x)\exp(-ix/\nu)dx,
\label{a1}
\end{equation}
where
\begin{eqnarray}
&& \varphi_2(x)= \frac{2\cosh
x}{\sinh^3x}-\frac{1}{x\sinh^2x}-\frac{\coth x}{x^2},
\nonumber \\
&& \varphi_3(x)=\left(\frac{2}{3}-\frac{1}{x^2}\right)\coth x+
\frac{1}{x\sinh^2x}. \label{a2}
\end{eqnarray}
The integrals in Eq.(\ref{a1}) can be evaluated closing the
integration contour in the lower half-plane and summing the residues
in the points $x=-in\pi$. Substituting the results into
Eq.(\ref{d1}) and then into Eq.(\ref{d5}) we obtain the pair
creation probabilities in an electric field in the case $r \ll
\nu^2/(1+\nu)$
\begin{eqnarray}
&& W_2=\frac{2\alpha
m^2r}{\omega}\left[\frac{1}{\nu^2(e^{\pi/\nu}-1)}-\frac{1}{\pi\nu}
\ln\left(1-e^{-\textstyle\frac{\pi}{\nu}}\right)\right],
\nonumber \\
&& W_3=\frac{2\alpha
m^2r}{\omega}\left[\frac{2}{3(e^{\pi/\nu}-1)}+\frac{2}{\pi^2}{\rm
Li}_2\left(e^{-\textstyle\frac{\pi}{\nu}}\right)-\frac{1}{\pi\nu}
\ln\left(1-e^{-\textstyle\frac{\pi}{\nu}}\right)\right], \label{a3}
\end{eqnarray}
where ${\rm Li}_2(x)=-\int\limits_0^x
\frac{\ln(1-t)}{t}dt=\sum\limits_{n=1}^{\infty}\frac{x^n}{n^2}$ is
the Euler dilogarithm.

In the case $\nu \ll 1$ one has
\begin{equation}
W_2=\frac{2\alpha
m^2r}{\omega\nu^2}e^{-\frac{\pi}{\nu}}\left(1+\frac{\nu}{\pi}\right),
\quad W_3=\frac{2\alpha
m^2r}{\omega\nu^2}e^{-\frac{\pi}{\nu}}\left(\frac{\nu}{\pi}+2\nu^2
\left(\frac{1}{3}+\frac{1}{\pi^2}\right)\right). \label{a4}
\end{equation}

At $\nu\ll 1$ the photon energy region $r \sim \nu^2$ remains
unexplained only. We close the integration contour in the lower
half-plane in Eq.(\ref{b0}) in the following way
\begin{equation}
{\rm Im}\Omega_i = i\frac{\alpha
m^2}{2\pi}\int\limits_{-1}^{1}dv\sum\limits_{n=1}^{\infty} \oint
f_i(v,x)\exp \left[i\psi(v,x)\right]dx,\label{a5}
\end{equation}
where the path of integration is any simple closed contour around
the points $-i\pi n$. Expanding the function entering in
Eq.(\ref{d3}) over variables $\xi_n=x+i\pi n~(|\xi_n|\sim
\sqrt{r}\sim \nu)$ and keeping the main terms of decomposition we
find
\begin{eqnarray}
&&f_2(v,x) \simeq -\frac{2}{\xi_n^3}\left[1+(-1)^{n+1}\cos
(vn\pi)\right]-f_3(v,x),\quad f_3(v,x) \simeq
(-1)^n\frac{iv}{\xi_n^2}\sin(vn\pi);
\nonumber \\
&& \psi(v,x) \simeq -\frac{2r}{\xi_n\nu}\left[1+(-1)^{n+1}\cos
(vn\pi)\right]-\frac{\xi_n}{\nu} +\frac{i\pi
n}{\nu}\left(1+r(1-v^2)\right)
\nonumber \\
&& +\frac{2ir}{\nu}(-1)^n v \sin(vn\pi)\label{a6}
\end{eqnarray}
Because of appearance of the factor $\exp(-\pi n/\nu)$, in the case
$\nu\ll 1$ the main contribution to the sum in Eq.(\ref{a5}) gives
the term $n=1$, then
\begin{eqnarray}
\hspace{-10mm}&&f_2(v,x)=-\frac{4\cos^2(v\pi/2)}{\xi^3}-f_3(v,x),\quad
\psi(v,x)=\frac{4r\cos^2(v\pi/2)}{\xi\nu}-\frac{\xi}{\nu}
\nonumber \\
\hspace{-10mm}&&  + \frac{i\pi}{\nu}\left(1+r(1-v^2)\right)
-\frac{2ir}{\nu} v \sin(v\pi),\quad
f_3(v,x)=-\frac{iv}{\xi^2}\sin(v\pi),\quad \xi=\xi_1. \label{a7}
\end{eqnarray}
Using the integrals Eq.(7.3.1) and Eq.(7.7.1)(11) in \cite{BE} and
substituting the result in Eq.(\ref{d1}) and then in Eq.(\ref{d5})
we find
\begin{eqnarray}
\hspace{-10mm}&&W_2=2\frac{\alpha
m^2}{\omega}e^{-\frac{\pi(1+r)}{\nu}}
\Bigg[I_1^2\left(\frac{2\sqrt{r}}{\nu}\right)-\frac{\nu}{\pi}
\left(I_0^2\left(\frac{2\sqrt{r}}{\nu}\right)-1\right)
\nonumber \\
\hspace{-10mm}&& +\frac{3\sqrt{r}}{\pi}
I_1\left(\frac{2\sqrt{r}}{\nu}\right)I_0\left(\frac{2\sqrt{r}}{\nu}\right)\Bigg],\quad
W_3 = \frac{\alpha m^2}{\omega}\frac{\nu}{\pi}
e^{-\frac{\pi(1+r)}{\nu}}
\left(I_0^2\left(\frac{2\sqrt{r}}{\nu}\right)-1\right),
 \label{a8}
\end{eqnarray}
where $I_n(z)$ is the Bessel function of imaginary argument. At
calculation of the correction terms $\propto \nu, \sqrt{r}$ the
integration by parts in the integral over $v$ was performed. The
found probability Eq.(\ref{a8}) is applicable for $r \leq \nu$. The
first term of decomposition of probabilities in Eq.(\ref{a8}) in
powers of $r$ coincide with Eq.(\ref{a4}) if in expression for $W_3$
the terms $\propto \nu^2$ are omitted.

For $r \gg \nu^2$  the asymptotic representation  $I_n(z)\simeq
e^z/\sqrt{2\pi z}$ can be  used. As a result one obtains the
probability Eq.(\ref{b8}).  If in Eq.(\ref{a8}) one omits the
correction terms $\propto \nu, \sqrt{r}, r/\nu$, than one gets the
result found in \cite{Vo} using completely different approach.

The curves in Fig.6 characterize the applicability of Eq.(\ref{a8})
for different values of parameter $\nu$ in the region of very low
energies. It is seen that at $r > \nu$ Eq.(\ref{a8}) is broken.

\begin{figure}[h]
  \centering
  \includegraphics[width=0.76\textwidth
  ]{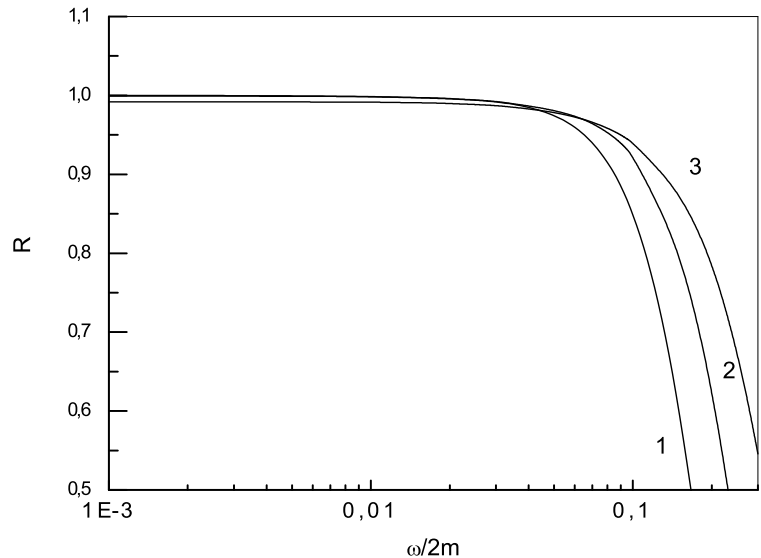}\hspace{0.025\textwidth}
 \caption{The ratio of the total probabilities of pair creation by a photon
averaged over the photon polarizations Eq.(\ref{d5a})
$R=W^{(l)}/W^{(ex)}$, where $W^{(l)}$ is given by Eqs.(\ref{a8}),
$W^{(ex})$ is given by Eq.(\ref{d5}), for $\nu=0.01$ (curve 1), for
$\nu=0.03$ (curve 2) and for $\nu=0.1$ (curve 3) vs $\omega/2m$.}
\label{Fig:6}
 \end{figure}

\section{Conclusion}

We considered the process of pair creation by a photon in an
electric field. The probability of the process is calculated using
the different approaches.  In the case $\nu \ll 1$ the standard
quasiclassical approximation is applicable for the energy parameter
$r \gg \nu^{-2/3}$. For $\nu=0.01$ the averaged over photon
polarization probability of pair creation $W$ Eq.(\ref{d5a})
coincides with the standard quasiclassic probability $W^{(SQA)}$
Eq.(\ref{c6}) within accuracy better than 1\% starting from $r=160$.
For $\nu=0.001$  $W$ coincides with $W^{(SQA)}$ within accuracy
better than 1\% starting from $r=700$ and  for $\nu=0.0001$ $W$
coincides with $W^{(SQA)}$  within accuracy better than 1\% starting
from $r=3200$. These estimates are in a good agreement with the
values of correction term in Eq.(\ref{b7}).  Note that always
$W>W^{(SQA)}$.

Similar situation was observed for the process of pair creation by a
photon in a magnetic field \cite{BK1} where the standard
quasiclassical approximation is applicable even in the case $\mu >
1$ at $r \gg \mu$. From Fig.2 it is seen that in an electric field
for $\nu=1$ the quasiclassical approximation is valid beginning from
values $r$ very close to 1.

The result is universal and the same in electric and magnetic fields
if $\nu=\mu$. However the corrections to SQA (see
Eqs.(\ref{c8})-(\ref{c10})) change sign at $\mu^2 \rightarrow
-\nu^2$.

It should be noted that the consideration based on the polarization
operator gives only the total probability of pair creation by a
polarized photon (real or virtual). The standard quasiclassical
method permits to obtain also the spectral, the angular
distributions as well as the polarization of the particles of
created pair (see \cite{BKS1}).

For lower values of $r$ the probability of pair creation in an
electric field is much higher than in a magnetic field. Besides the
pair creation in an electric field is possible also at $r<1$
(although the probability is exponentially small). This phenomenon
can be interpreted that an electric field help to draw out the pair
from the vacuum.  At $r\ll 1(\omega\ll m)$ the inverse situation
occurs: a photon helps to electric field to do this work (the factor
$\exp(4\sqrt{r}/\nu)$ Eq.(\ref{a4})) and for $r \ll \nu^2$ this aid
becomes negligible.

The above analysis is not complete if the probability of direct pair
creation by an electric field (vacuum probability) is essential.
Then Eq.(\ref{d5}) gives the partial contribution to the probability
under consideration and defines the photon lifetime. At $\nu\ll
1~(E\ll E_0)$ the vacuum contribution is negligible. But even in
this case for very low photon energies in the region $r\sim
\nu^2~(\omega\sim eE/m)$ the probability Eq.(\ref{a4}) becomes
comparable with the vacuum probability.

For lower energies $r \ll \nu^2$ Eq.(\ref{a4}) defines the
probability of photon absorption by the particles of created by an
electric field.

\end{document}